\documentclass[twoside,12pt]{article}
\usepackage{epsf,epsfig,amssymb,amsmath,graphicx,float}
\usepackage{color}

\usepackage{hyperref}
\usepackage{amsmath}
\usepackage{indentfirst}

\hypersetup{colorlinks,citecolor=blue,linkcolor=red,urlcolor=red}

\begin{document}
\renewcommand{\thefootnote}{\fnsymbol{footnote}}
\begin{titlepage}
\begin{center}		
\vspace{1cm}			
{\Large {\bf Relic abundance of dark matter with coannihilation in non-standard
    cosmological scenarios }}
\vspace{1cm}

{\bf Fangyu Liu, Hoernisa Iminniyaz}\footnote{Corresponding author, email:wrns@xju.edu.cn}

\vskip 0.15in
{\it
	%$^a$
	{School of Physics Science and Technology, Xinjiang University, \\
		Urumqi 830017, China} \\
}

\begin{abstract}		
  We investigate the relic abundance of dark matter from coannihilation in
  non-standard cosmological scenarios. We explore the effect of coannihilation
  on the relic density of dark matter and freeze
  out temperature in quintessence model with kination phase
  and brane world cosmological scenarios. Since the Hubble expansion
  rate is enhanced in quintessence and brane
  world cosmological models, it causes the larger relic density compared to
  that in the standard one. On the other hand, the relic density of dark
  matter is
  decreased due to the coannihilation in the standard cosmological scenario.
  After including coannihilation in quintessence or brane world cosmological
  scenarios, we find the decrease
  of the relic density of dark matter is slightly slower
  than that in the standard cosmological scenario. 
\end{abstract}
\end{center}
\end{titlepage}
\setcounter{footnote}{0}

\section {Introduction}
 Recent Planck results provide the abundance of Dark Matter (DM)
  as $\Omega_{\rm DM}h^2=0.120\pm0.001$ \cite{Planck:2018vyg} . Although the
  precise value of DM abundance is known from the observation, 
  the nature of DM is still a mystery for the scientists. The Weakly
  Interacting Massive Particles (WIMPs) are the most popular
  candidates for DM. The best motivated one is the lightest
  supersymmetric particle (LSP) neutralino which is stabalized
  due to R--Parity, we denote it by $\chi_1$. It is usually assumed that WIMPs
  were in full thermal
  and chemical equilibrium when the interaction rate of DM particles $\Gamma$
  is greater than the expansion rate $H$ of the universe. WIMPs freeze out
  when $\Gamma \textless H$.
    
  The Big-Bang nucleosynthesis (BBN) successfully predicted the abundances of 
  light element isotopes D, $^{3}$He, $^{4}$He, and $^{7}$Li, and it is known
  that the
  nucleosynthesis takes place at the temperature scale $\sim 1$ Mev
  \cite{ParticleDataGroup:2018ovx}. On the other hand, we don't have
  observational
  information of the universe prior to BBN. Therefore, it is necessary to
  consider scenarios that could modify the cosmic expansion rate in the
  period before BBN and some of which yielded significant physical
  implications. In non-standard cosmological scenarios, the expansion rate of
  the universe is larger or smaller than the standard one. 
  The evolution of number density $n$ of DM is described by the Boltzmann
  equation \cite{standard-cos}. Solving the Boltzmann equation, we find that
  the non-standard expansion rate leaves its imprint on the relic density of DM
  \cite{Allahverdi:2020bys,Berlin:2016vnh}. In order not to spoil the
  predictions of BBN, the non-standard cosmic expansion rate must returned to
  the form of the radiation-dominated era at the beginning of BBN.
  
  There are several kinds of modified cosmological scenarios. One
    example is the
  quintessence model with kination phase. If the universe experience a period
  called ``kination", in which the energy density $\rho(a)$ of the universe is
  dominated by the kinetic energy of the scalar field with $a$ being the scale
  factor, then $\rho(a)$ drops very quickly as $1/a^6$
  \cite{Weinberg:2008zzc,Joyce:1996cp,Salati:2002md,DEramo:2017gpl,AristizabalSierra:2023bah,Co:2021lkc}. 
  The abundance of DM can be significantly affected by this kind of fast
  evolution. \cite{Salati:2002md,DEramo:2017gpl}
  discussed the consequences of
  kination period on the relic abundance of DM. They found the resultant relic
  density is increased due to the enhanced Hubble expansion rate in kination
  model. Another
  example of the modified cosmology is the brane world
  scenario which also predicted the Hubble expansion rate of the
  universe is different from the standard cosmology
  \cite{Langlois:2002bb,Randall:1999vf,Binetruy:1999hy,Ichiki:2003hf,Binetruy:1999ut}. In brane world scenario,
  the ordinary matter is assumed to be confined onto a three-dimensional
  subspace, called brane which is embedded in a higher dimensional spacetime,
  named bulk. Brane world scenario is relatively different from the standard
  cosmology of four-dimensional universe. Extra-dimension effects may play an
  important role in the early universe.  the DM relic density is also
  affected by the physics of extra
  dimensions. \cite{Okada:2004nc,AbouElDahab:2006glf} explored the impact of
  brane cosmology on the DM relic density. They conclude the relic density is
  considerably enhanced compared to that in the standard cosmology.

  The standard calculation of relic density of DM included the annihilations of
  LSP only. 
  When the next-to-the-lightest
  supersymmetric particles (NLSP) are slightly heavier than the LSP, the relic
  abundances of LSP is determined both by the annihilation cross section
  of LSP and the annihilation of the heavier particles. The heavier
  particles later decay into the LSP \cite{Griest:1990kh}. This process is
  called ``coannihilation''.
  In \cite{Griest:1990kh}, the authors considered the case where the
  squark's mass is slightly heavier than the LSP in the standard cosmological
  scenario. They found the relic abundance of
  the LSP is significantly reduced because of the coannihilation.
  In our work, we investigate the relic density of LSP with
  coannihilation in the non-standard cosmological scenarios, which includes
  quintessence model with kination phase and brane world cosmology. In these
  two non-standard cosmological models, we
  plot the ratio of the relic density of DM without coannihilation to the case
  including coannihilation as a function of the
  relative mass splitting for different modifications and cross
  section enhancements, here the relative mass spilitting is the ratio
  of the mass difference between 
  NLSP and LSP to the mass of LSP. For comparison we also plot the ratio of
  the relic density as a function of the
  relative mass splitting in the standard cosmological scenario to the
  non-standard cosmological scenarios which includes coannihilation. 
  The enhanced Hubble
  expansion rate leads to the earlier particle freeze out and increased relic
  density in kination and brane world cosmological scenarios. On
  the other hand, the coannihilations reduce the relic density of DM. After
  considering the coannihilation in the non-standard cosmology,
  we found the decrease of the relic density in those models is slower than
  the standard cosmological scenario.
  The effect of coannihilation on the relic density depends on the size of the
  modification factor in non-standard cosmology. 
   For larger modification, the increase of the DM relic density is
 sizable in kination model and brane world cosmology,
 therefore, after including coannihilation, there is slight decrease of the
 relic abundance in
 contrast with the minor modification of Hubble rate.
 In other words, compared to the increase of
 the relic density because of the enhanced Hubble expansion rate, the decrease
 of the relic density due to coannihilation of DM is insignificant for
 larger modification. We found the constraints on the
 Hubble enhancement factors and the relative mass splitting which let the 
 coannihilation has effect on the relic density of DM.
 
We organize the paper as follows. In section 2, we briefly review the
quintessence model with kination phase and brane world cosmology. In
  section 3, we discuss the coannihilation in the standard cosmological
  scenario and then extend it to the non-standard cosmological scenarios.
  We investigate the effect of coannihilation on the relic abundance in detail
  in quintessence model with kination phase and brane world cosmology. The last section is devoted to the conclusion and summary.

  \section{Expansion rate of the universe in kination  and brane world
    cosmology}
  The expansion rate of the universe is given by Friedmann equation 
\begin{equation}\label{eq5}
  \Biggl(\frac{{\rm d}a}{{\rm d}t}\Biggr)^{2}+k=\frac{8\pi G\rho a^{2}}{3}\,,
\end{equation}
where $G$ is the Newton’s gravitational constant. The contribution of
non-relativistic and
  relativistic matter to quantity $\rho a^{2}$ grows as ``$1/a$'' and
  ``$1/a^2$'' respectively. At sufficiently early times,
  $a\rightarrow0$, the constant ``$k$'' can be neglected\cite{Weinberg:2008zzc}. Then, Eq.(\ref{eq5}) becomes
\begin{equation}\label{eq6}
  H^2=\frac{8\pi G \rho}{3}\,,
\end{equation}
here $H = \dot{a}/a$. In the standard cosmological scenario, the dominant
radiation energy density $\rho_{\rm rad}$ is
\begin{equation}\label{eq7}
	\rho_{\rm rad}=\frac{\pi^2}{30}g_{*}T^4\,.
\end{equation}
Here $g_*$ is the effective number of relativistic degrees of freedom. Therefore, the standard expansion rate of the universe is
\begin{equation}
H_{\rm std}=\frac{2\pi T^2}{m_{\rm Pl}}\sqrt{\frac{\pi g_*}{45}}
\end{equation}
where $m_{\rm Pl}\equiv1/\sqrt{G}=1.22\times10^{19}$ GeV is Planck mass.
  
  Quintessence is a time-varying vacuum energy that depends on one or more
  scalar fields \cite{Weinberg:2008zzc}. Considering a single real scalar
  field, its action is
\begin{equation}\label{eq1}
  I_{\varphi}=-\int d^{4}x\sqrt{-Detg}\Biggr[\frac{1}{2}g^{ \mu\nu}\frac{\partial\varphi}{\partial x^{ \mu}}\frac{\partial\varphi}{\partial x^{\nu}}+V(\varphi)\Biggr]\,,
\end{equation}
 where $g^{\mu\nu}$ is Robertson-Walker metric and $\varphi$ depends only on
  the time due to the homogeneity. Varying the field $\delta\varphi(t)$, the
  action is unchanged $\delta I_{\varphi}=0$, then one can get the field
  equation
\begin{equation}\label{eq2}
  \frac{{\rm d^2}\varphi(t)}{{\rm d}t^2}+3H\frac{{\rm d}\varphi(t)}{{\rm d}t}+\frac{{\rm d}V(\varphi)}{\rm  d\varphi}=0\, .
\end{equation}
The energy density $\rho_{\varphi}=\frac{1}{2}\dot{\varphi}^{2}+V$ is derived
from Eq.(\ref{eq2}) as
\begin{equation}\label{eq3}
  \rho_{\varphi}=\rho_{\varphi}(t_0)e^{-\int_{t_0}^{t}\frac{6}{1+\beta(t)}H(t)dt}=\rho(t_0)e^{-\int_{a_0}^{a}\frac{6}{1+\beta(a)}\frac{da}{a}}\,,
\end{equation}
where $\beta$ is defined as
$\beta\equiv V(\varphi)/(\frac{1}{2}\dot{\varphi}^{2})$ \cite{Joyce:1996cp} .
The energy density of a special period called ``kination'' is obtained from Eq.\eqref{eq3}. During ``kination'' period, the kinetic energy  $\frac{1}{2}\dot{\varphi}^{2}$ dominates over the potential
energy $V(\varphi)$ i.e. $\frac{1}{2}\dot{\varphi}^{2}\gg V$ and
$\beta\rightarrow0$. At kination phase, performing the integration in
Eq.(\ref{eq3}), then $\rho_{\varphi}$ is obtained as \cite{Joyce:1996cp}
\begin{equation}\label{rho_varphi}
  \rho_{\varphi}\propto\frac{1}{a^{6}}\,.
\end{equation}

  The expansion rate in kination model is constrained
  by BBN \cite{Salati:2002md}, which means the expansion rate in kination model
  returns to the standard case before BBN$(\sim 1 $ MeV) starts. 
%  In other words, the radiation energy will become dominant.
  Defining $\eta$ as
  \begin{equation}\label{Tr definition}
  	\eta =\frac{\rho_\varphi(T_{\rm r})}{\rho_{\rm rad}(T_{\rm r})}\, ,
  \end{equation}
  we choose $T_{\rm r}=10$ MeV and when $\eta\ll 1$, the universe is
  radiation dominated. 
  Combining Eq.(\ref{eq7}) and Eq.(\ref{Tr definition}) with
Eq.(\ref{rho_varphi}), again using the entropy conservation 
\begin{equation}
  sa^3={\rm constant} \quad i.e. \quad a\propto\frac{1}{Tg_{\rm *s}^{1/3}}\,,
\end{equation}
where $g_{\rm *s}$ is the effective number of entropic degrees of freedom,
then the expression of the energy density $\rho_{\varphi}$ is written as \cite{Joyce:1996cp, Salati:2002md, DEramo:2017gpl} 
\begin{equation}\label{rhophi}
  \rho_\varphi=\eta \, \rho_{\rm rad}(T_{\rm r})\left[  \frac{g_{\rm *s}(T)}{g_{\rm *s}(T_{\rm r})}\right]
  ^2\left( \frac{T}{T_{\rm r}}\right) ^6\,.
\end{equation}
  Now, we can have the relationship between the expansion rates in the
  standard cosmology and kination model,
\begin{equation}\label{eq9}
  H_{\rm que}^{2}=\frac{8\pi G}{3}\rho_{\rm rad}(1+\frac{\rho_{\varphi}}{\rho_{\rm rad}})\,.
\end{equation}
Inserting Eq.(\ref{eq7}) and Eq.(\ref{rhophi}) into Eq.(\ref{eq9}), one can
obtain
\begin{equation}\label{H_{kin}}
  \frac{H_{\rm que}}{H_{\rm std}}=\sqrt{1+\eta\,\frac{g_*(T_{\rm r})}{g_*(T)}\left[  \frac{g_{\rm *s}(T)}{g_{\rm *s}(T_{\rm r})}\right]  ^2\left( \frac{T}{T_{\rm r}}\right) ^2}\,.
\end{equation}
We use the dimensionless quantity $x = m_1/T$ with $m_1$ being the mass of the
DM particle $\chi_1$, then 
\begin{equation}\label{H_{kin}}
  \frac{H_{\rm que}}{H_{\rm std}}=\sqrt{1+\eta\,\frac{g_*(x_{\rm r})}{g_*(x)}\left[  \frac{g_{\rm *s}(x)}{g_{\rm *s}(x_{\rm r})}\right]  ^2\left( \frac{x_{\rm r}}{x}\right) ^2}\,.
\end{equation}

  In
  Ref.\cite{Langlois:2002bb,Randall:1999vf,Binetruy:1999hy,Ichiki:2003hf,Binetruy:1999ut},
  authors provided comprehensive introduction to brane cosmology. In brane world cosmology, standard model particles are confined on a
  three-dimensional subspace ``3-brane'', embedded in a higher dimensional 
spacetime. RS II model is the simple and interesting brane world model
    which is proposed by Randall and Sundrum (RS) \cite{Randall:1999vf}. 
  In this model, the 4-dimensional universe is realized on the 3-brane with a
  positive tension, located at the ultra-violet boundary of the five
  dimensional Anti de-Sitter spacetime. In this framework, the expansion rate 
  is given by the modified Friedmann equation, 
\begin{equation}
  H_{\rm bra}^2=\frac{8\pi G}{3}\rho_{\rm rad}(1+\frac{\rho_{\rm rad}}{2 \lambda})\,,
\end{equation}
where $\lambda$ is the brane tension
\begin{equation}
  \lambda=\frac{48\pi M_5^6}{m_{\rm Pl}^2}\,,
\end{equation}
  here $M_5$ is the five-dimensional Plank mass. In the brane
  world cosmological scenario, the radiation energy density becomes dominant
  at BBN ($\sim$1 Mev) too,
\begin{equation}
%  \rho_c=\frac{96\pi M_5^6}{m_{\rm Pl}^2}\gg\rho=\frac{\pi^2}{30}g_*T^4\,,
%  \qquad T= 1 {\rm MeV}\,,
    \frac{\rho_{\rm rad}}{2 \lambda}\ll 1,\ \ {\rm when}
\ \ T= 1\, {\rm MeV}\,,
\end{equation}
  i.e. $M_5\gg1.1\times10^4$ GeV, where $g_*({\rm 1\, MeV})=10$. In addition, the
  precision measurements of the gravitational law in submillimeter range give
  the further limit on $M_5>10^8$ GeV \cite{Langlois:2002bb}. In this
  work, we consider the constraint by BBN. Therefore, the relationship between
  the expansion rates in standard cosmology and brane world cosmological
  model is
  % %
  %
\begin{equation} \label{H_{bra}}
  \frac{H_{\rm bra}}{H_{\rm
      std}}=\sqrt{1+\frac{\rho_{\rm rad}}{\rho_0}}=
  \sqrt{1+\frac{\pi g_*m_{\rm Pl}^2m_1^4}{2880M_5^6x^4}}\,.
\end{equation}

\section {Relic density with coannihilation in non-standard cosmology}
In Refs.\cite{Griest:1990kh,Baker:2015qna,Edsjo:1997bg}, authors provided
detailed analysis to coannihilaiton. Here we review the
Boltzmann equation obtained in Ref.\cite{Griest:1990kh} including the
coannihilation very briefly. For a series of supersymmetric particles $\chi_{\rm i}$ with
increasing mass
$m_1 \textless \dotsm m_{\rm i}\textless \dotsm  m_{\rm j}\textless \dotsm m_{\rm N}$, with
internal degrees of freedom $g_1,\dotsm g_{\rm i},\dotsm g_{\rm j},\dotsm g_{\rm N}$, the Boltzmann equation for the total number densities $n=\sum_{\rm i=1}^{N}n_{\rm i}$ of
particles $\chi_{\rm i}$ is 
\begin{equation}\label{c b e}
  \frac{{\rm d}n}{{\rm d}t}=-3Hn-\sum_{\rm i,j=1}^{N}\langle\sigma_{\rm ij}v\rangle(n_{\rm i}n_{\rm j}-n_{\rm i}^{\rm eq}
  n_{\rm j}^{\rm eq})\,,
\end{equation}
where $\langle\sigma_{\rm ij}v\rangle$ is the thermal average of the pair
annihilation cross sections of the particles $\chi_{\rm i}$ and $\chi_{\rm j}$ times
their relative velocity, $n^{\rm eq}_{\rm i}$ is the equilibrium value of the number
density. Since the scattering rate of supersymmetric particles off particles
in the thermal background is much faster than their annihilation rate, then
$n_{\rm i}/n$ is well
approximated by its equilibrium value $n^{\rm eq}_{\rm i}/n^{\rm eq}$, i.e. $n_{\rm i}/n\approx n_{\rm i}^{\rm eq}/n^{\rm eq}$, where $n_{\rm i}^{\rm eq}\approx g_{\rm i}(m_{\rm i}T/2\pi)^{3/2}{\rm exp}(-m_{\rm i}/T)$.
Defining
  \begin{equation}
r_{\rm i}\equiv\frac{n_{\rm i}^{\rm eq}}{n^{\rm eq}}=\frac{g_{\rm i}(1+\Delta_{\rm i})^{3/2}\exp(-x\Delta_{\rm i})}{g_{\rm eff}}\:,
  \end{equation} 
   where $\Delta_{\rm i}=(m_{\rm i}-m_1)/m_1$, and
   \begin{equation}\label{geff}
   	g_{\rm eff}=\sum_{\rm i=1}^{N}g_{\rm i}(1+\Delta_{\rm i})^{3/2}\exp(-x\Delta_{\rm i})\, ,
   \end{equation}
  then Eq.(\ref{c b e}) transforms into the standard form of the Boltzmann equation

\begin{equation}\label{Boltz}
  \frac{{\rm d}n}{{\rm d}t}=-3Hn-\langle\sigma_{\rm eff}v\rangle(n^2-n_{\rm eq}^2)\,,
\end{equation}
where
\begin{equation}\label{sigmaeff}
  \sigma_{\rm eff}=\sum_{\rm ij}^{N}\sigma_{\rm ij}r_{\rm i}r_{\rm j}=\sum_{\rm ij}^{N}\sigma_{\rm ij}\frac{g_{\rm i}g_{\rm j}}{g_{\rm eff}^2}(1+\Delta_{\rm i})^{3/2}(1+\Delta_{\rm j})^{3/2}\exp[-x(\Delta_{\rm i}
  + \Delta_{\rm j})]\,.  
\end{equation}

  Now, we analyze the relic density of DM with coannihilation in non-standard
  cosmology. For convenience, a useful definition is applied as $Y\equiv n/s$. In kination and brane world models, using Eqs.(\ref{H_{kin}}),
  (\ref{H_{bra}}) and $t=0.301g_{\ast}^{-1/2}\,m_{\rm Pl}\,x^2/m^2$, the Boltzmann Equation (\ref{Boltz}) is written as 
\begin{equation}\label{Boltzmann Equation}
  \frac{{\rm d}Y}{{\rm d}x}=-\frac{2\pi^2}{45}\frac{g_{\rm \ast s}m_1^{\,3}x^{-4}}{H_{\rm nstd}} \langle\sigma_{\rm eff}v\rangle(Y^2-Y_{\rm eq}^2)\,,
\end{equation}
where 
\begin{equation}
  Y_{\rm eq}=\sum_{\rm i=1}^{N}\frac{n_{\rm i}^{\rm eq}}{s}=0.145\frac{x^{3/2}e^{-x}}{g_{\rm \ast s}}\Biggl[\sum_{\rm i=1}^{\rm N}g_{\rm i}(1+\Delta_{\rm i})^{3/2}\exp(-x\Delta_{\rm i})\Biggr]\,.
\end{equation}
Here $H_{\rm nstd}$ is the Hubble rate in the non-standard cosmological
scenarios which includes quintessence model with kination phase and brane world
cosmology. We solve the Boltzmann equation (\ref{Boltzmann Equation}) by
following the standard picture of the DM
particle evolution \cite{Scherrer:1985zt}. DM particles are in thermal
equilibrium at high temperature in the early universe. When the temperature
decreases, the equilibrium number densities of the DM particles drops
exponentially. In the end, the interaction rate becomes smaller than the
expansion rate, then the DM particles decoupled from the equilibrium state and
their number densities become almost constant from that freeze out
point. Usually, the non-standard cosmic expansion rate is greater
than the standard cosmic expansion rate prior to BBN. It leads the DM
particles decouple earlier than in the standard cosmology. 

While the Boltzmann equation (\ref{Boltzmann Equation}) can be computed
numerically, it is still useful to obtain the analytic solution for
Eq.(\ref{Boltzmann Equation}). We rewrite it in terms of $\delta = Y-Y_{\rm eq} $,
\begin{equation}\label{delta}
  \frac{\rm d \delta}{{\rm d}x}=-\frac{{\rm d}Y_{\rm eq}}{{\rm d}x}-\frac{2\pi^2}{45}\frac{g_{\rm \ast
      s}m_1^{\,3}x^{-4}}{H_{\rm nstd}}
  \langle\sigma_{\rm eff}v\rangle\,\delta[\delta+2Y_{\rm eq}]\,.
\end{equation}
The solution of equation (\ref{delta}) is considered in two
regimes. $Y$ tracks its equilibrium value at high temperature, in this
situation, $\delta^2$ and ${\rm d }\delta/{\rm d}x$ are negligible. The Boltzmann equation
is simplified to
\begin{equation}\label{delta-simp}
  \frac{{\rm d}Y_{\rm eq}}{{\rm d}x}=-\frac{4\pi^2}{45}\frac{g_{\rm \ast
      s}m_1^{\,3}x^{-4}}{H_{\rm nstd}}
  \langle\sigma_{\rm eff}v\rangle\,\delta\,Y_{\rm eq}\,.
\end{equation}
The approximate solution of Eq.(\ref{delta-simp}) is
\begin{equation}\label{Y-Y_{eq}}
  \delta \approx\frac{g_*^{1/2}x^2H_{\rm nstd}}{0.528g_{\rm *s}m_1m_{\rm Pl}\langle\sigma_{\rm eff}v\rangle H_{\rm std} }\,,
\end{equation}
where ${\rm d}Y_{\rm eq}/{\rm d}x\approx-Y_{\rm eq}$ is applied for $x\gg 1$. This solution for
$\delta$ is used to fix the scaled freeze out temperature $x_{f}$.

At late times ($x\gg x_f$), $\delta \approx Y\gg Y_{\rm eq}$, therefore
$Y_{\rm eq}$ and ${\rm d}Y_{\rm eq}/{\rm d}x$ can be neglected, then Eq.(\ref{delta}) becomes
\begin{equation}\label{delta-simp1}
  \frac{\rm d \delta}{{\rm d}x}=-\frac{2\pi^2}{45}\frac{g_{\rm \ast
      s}m_1^{\,3}x^{-4}}{H_{\rm nstd}}
  \langle\sigma_{\rm eff}v\rangle\,\delta^2 \,.
\end{equation}
Assuming $\delta(x_f)\gg \delta(\infty)$, an approximate
solution is obtained by integrating Eq.(\ref{delta-simp1}) from $x_f$ to
$\infty$,
\begin{equation}\label{Yinfty}
  Y(\infty)=\frac{g_*^{1/2}}{0.264g_{\rm *s}m_{\rm Pl}m_1}\frac{1}{\int_{x_f}^{\infty}(H_{\rm std}/H_{\rm nstd})\,\langle\sigma_{\rm eff}v\rangle/x^2{\rm d}x}\,.
\end{equation}
The modified expansion rate and coannihilation both affect the scaled freeze out
temperature $x_f$. Freeze out occurs when $\delta$ is of the same order as the
equilibrium value of $Y$,
\begin{equation}\label{xf-c}
  \delta(x_f)=c\;Y_{\rm eq}(x_f)\,,
\end{equation}
where $c$ is a constant. When $c=\sqrt{2}-1$, the approximate result matches
with the numerical one well \cite{Scherrer:1985zt}. 
Substituting Eq.(\ref{Y-Y_{eq}}) into Eq.(\ref{xf-c}), the scaled freeze out
temperature $x_f$ including coannihilation with the modified expansion rate is
obtained
\begin{equation}
  x_f=\ln\left( \frac{0.076\,c\,g_{\rm eff}\,m_1m_{\rm Pl}\langle\sigma_{\rm eff} v\rangle}{g_*^{1/2}x^{1/2}}\frac{H_{\rm std}}{H_{\rm nstd}}\right) \Bigg|_{x=x_f} \,,
\end{equation}

Considering a certain DM candidate and its pair annihilation cross section
$\langle\sigma  v\rangle$, the modified mechanism
``$H_{\rm nstd} + coannihilation$''  brings a significant difference
on the relic abundance compared to that in the standard mechanism
``$H_{\rm std}+pair\, annihilation$''. The differences are as follows,
\begin{subequations}
	\begin{center}
		\begin{align}
			&\Omega_{\rm nstd+co} = \Omega_{\rm std} \,
			\frac{\int_{x_{f,\,{\rm std}}}^{\infty} \langle\sigma v\rangle/x^2
				{\rm d}x}{\int_{x_{f,\,{\rm nstd+co}}}^{\infty}(H_{\rm std}/H_{\rm nstd}) \, \,
				\langle\sigma_{\rm eff}v\rangle/x^2 {\rm d}x} \,,\label{xf} \\		
			&x_{f,\,{\rm nstd+co}}= x_{f,\,\rm std} + \ln\left(\frac{x_{ f,\,{\rm std}}^{1/2}}{x_{f,\,{\rm nstd+co}}^{1/2}}\frac{H_{\rm std}}{H_{\rm nstd}} \frac{g_{\rm eff}} {g_1}\frac{\langle\sigma_{\rm eff} v\rangle}{\langle\sigma v\rangle _{\lvert _{x=x_{f,\,{\rm std}}} } }\right) \Bigg|_{x=x_{f,\,{\rm nstd+co}}} \,, \label{xf-xfstd} 
		\end{align}
	\end{center}
\end{subequations}
where $\Omega$ is the present-day mass density divided by the
critical density,  $\Omega \equiv \rho_{\chi}/\rho_c$ with  
$\rho_{\chi} =n m_1 = s_0 Y_{\chi}  m_1$ and the
critic density $\rho_c = 3 H^2_0 m^2_{\rm Pl}/8\pi$, 
where $s_0 \simeq 2900$ cm$^{-3}$ is the present entropy density and $H_0$ is
the Hubble constant,  $x_{f,\,\rm std}$ is the scaled freeze out temperature
in standard cosmology with pair annihilation. Here
\begin{equation}
	\Omega_{\rm std} h^2=\frac{s_0m_1h^2}{\rm \rho_c}Y_{\rm std}(\infty)
\end{equation}
and \begin{equation}
Y_{\rm std}(\infty)=\frac{g_*^{1/2}}{0.264g_{\rm *s}m_{\rm Pl}m_1}\frac{1}{\int_{x_{f,\,\rm std}}^{\infty}\langle\sigma v\rangle/x^2{\rm d}x}\,.
\end{equation}
here $ h = 0.674 \pm 0.005 $ is the scaled Hubble constant in units of 
$100$ km s$^{-1}$ Mpc$^{-1}$ \cite{Planck:2018vyg}.

Due to the presence of the non-standard expansion rates and coannihilation,
the relic density of DM changes significantly compared to that in the standard
case. We
follow the example taken by \cite{Griest:1990kh}, for
the case of two particles system, neutralino $\tilde{\chi}$ (denoted by $\chi_1$ ) and
squark $\tilde{q}$ (denoted by $\chi_2$), since
\begin{equation}
  \sigma_{22}(\tilde{q}\bar{\tilde{q}}\rightarrow \textsl{g}\textsl{g} )\approx(\alpha_s/\alpha)\sigma_{12}(\tilde{\chi}\tilde{q}\rightarrow q\textsl{g})\approx (\alpha_s/\alpha)^2\sigma_{11}(\tilde{\chi}\tilde{\chi}\rightarrow q\bar{q})\,,
\end{equation}
 Eqs.\eqref{sigmaeff} and \eqref{geff} become 
\begin{equation}
  \sigma_{\rm eff}=\sigma_{11}\left[\frac{1+Aw}{1+w}\right]^2\,, \qquad g_{\rm eff}=g_1(1+w)\,,
\end{equation} 
  where $w=(1+\Delta)^{3/2}\exp(-x\Delta)g_2/g_1$, $A= \alpha_s/\alpha$, $\Delta=(m_2-m_1)/m_1$, 
  ``$\textsl{g},\: q $'' denote a gluon and quark respectively,
  ``$\alpha_s,\:   \alpha$'' are the strong-interaction coupling and the
  electroweak coupling respectively.

\begin{figure}[H] 
	\begin{center}
		\hspace*{-0.5cm} \includegraphics*[width=7.35cm]{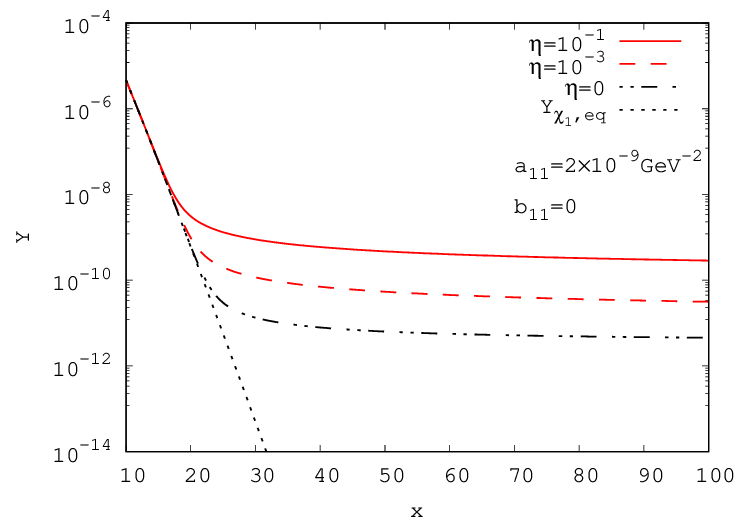}
		\put(-115,-12){(a)}
		\vspace{0cm}
		\hspace*{-0.5cm} \includegraphics*[width=7.35cm]{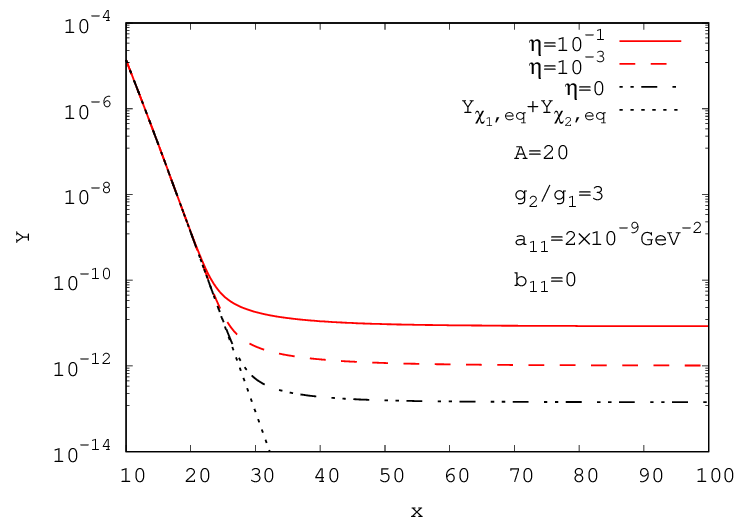}
		\put(-115,-12){(b)}
                \vspace{0cm}
                \hspace*{-0.5cm} \includegraphics*[width=7.35cm]{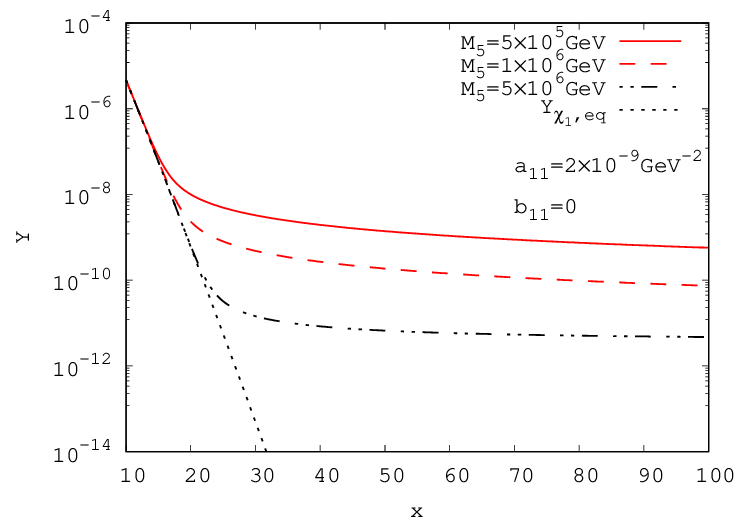}
		\put(-115,-12){(c)}
                \hspace*{-0.5cm} \includegraphics*[width=7.35cm]{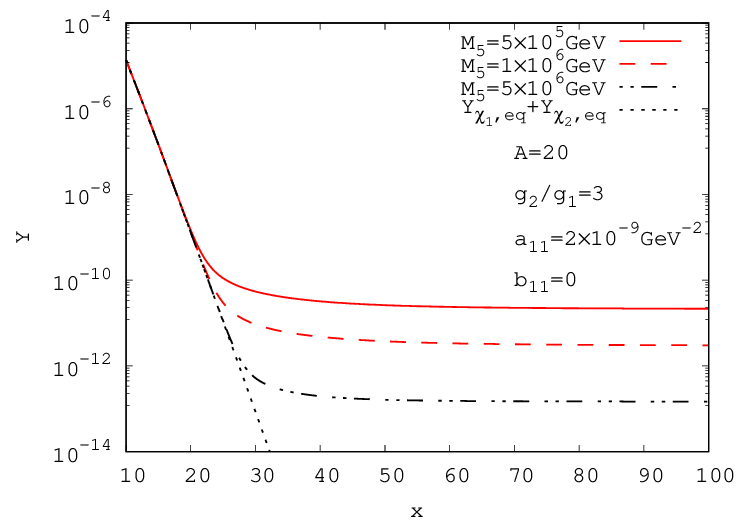}
		\put(-115,-12){(d)}
		\caption{\label{fig:z} \footnotesize
            Evolution of the relic abundance $Y$ of DM in
                  quintessence model with kination
			phase and brane world cosmology 
			 as a function of the inverse-scaled temperature
                         $x$. (a) and (c) are plotted without coannihilation;
                         (b) and (d) are with coannihilation.
			Here $m_1 = 100$ GeV, $g_{1} = 2$, $g_* = 90$.}  
	\end{center}
\end{figure}
Following, we discuss the effect of modified expansion rate on the relic
density of DM particle with coannihilation. The relic
           abundance $Y$ of DM with and without coannihilation as
            a function of the
        inverse-scaled temperature $x$ in quintessence model with
        kination phase and brane world cosmology is plotted   
in Fig.\ref{fig:z}. We set $g_*= g_{\rm *s}= {\rm constant}$ for
simplicity and take $m_1=100$ GeV, $T_{\rm r}=10$ MeV,
$g_*=g_{\rm *s}=90$, $A=20$, $g_2/g_1=3$, $x_{f,\,\rm std}=22$. Panels (a) and
(c) are for the case which is not including 
coannihilation in kination model and brane world cosmology; (b) and (d) are
plotted with coannihilation. We found the enhanced Hubble expansion rate in
kination model and brane world cosmology accelerates the decoupling of DM particles and
results larger values for the relic density. The size of increase depends on the
modification factor
$\eta$ in kination model and the five-dimensional Planck mass $M_5$ in brane
world cosmology. If we compare panel (a) with panel (b), we learn that the DM
relic density is reduced considerably after including coannihilation. The same
result is obtained for the case of brane world cosmology as shown in panels (c)
and (d). When $\eta = 0$ and
$M_5 = 5\times 10^6$ GeV, the standard case is recovered.   

\begin{figure}[H] 
	\begin{center}
		\hspace*{-0.5cm} \includegraphics*[width=8cm]{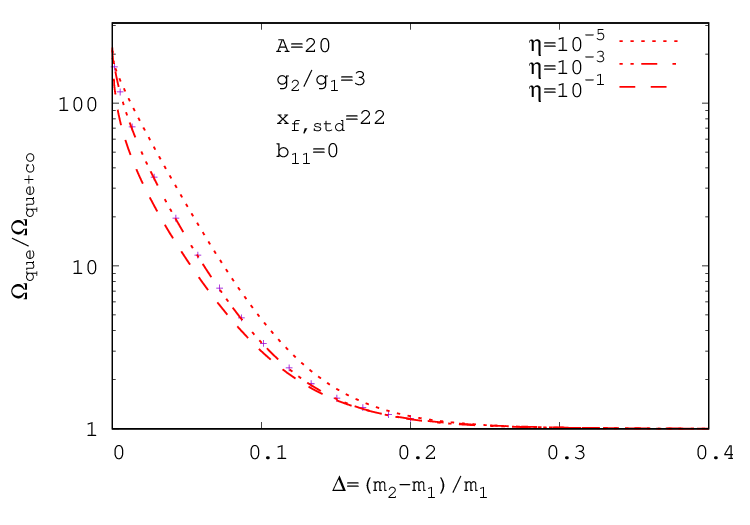}
		\put(-115,-12){(a)}
		\vspace{0.5cm}
		\hspace*{-0.5cm} \includegraphics*[width=8cm]{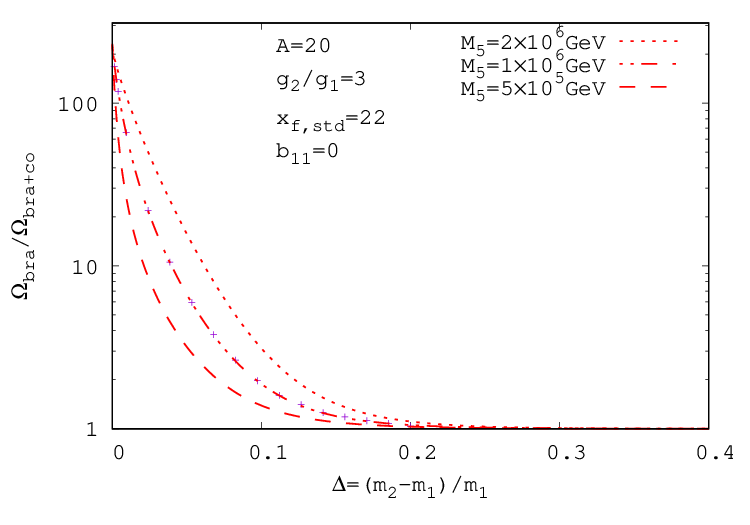}
		\put(-115,-12){(b)}
		\caption{\label{fig:a} \footnotesize
			The ratio of the relic density of DM including only
                        neutralino pair annihilation $\sigma_{11}$ to the case
                        including coannihilation
                        $\sigma_{\rm eff}$ as a
			function of the relative mass spliting $\Delta$ in
                        kination model and brane world cosmology.
			Here $m_1 = 100$ GeV, $g_{1} = 2$, $g_* = 90$.}  
	\end{center}
\end{figure}
Fig.\ref{fig:a} shows the ratio of the relic density of DM as a function of
relative mass splitting $\Delta$
for different modifications in
kination model and brane world cosmology without coannihilation to
the case including coannihilation.
Panel (a) is for the constant cross section
($s-$wave) which corresponds to the freeze out temperature
$x_{f,\rm\, std} = 22$ in kination model; panel (b) is
for the brane world cosmology. Here the line with plus sign is plotted using
the numerical solution of Eq.(\ref{Boltzmann Equation}). This figure shows the
extent of the coannihilation effect on the relic density in non-standard
cosmology. The Hubble expansion
rate is enhanced in kination and brane world cosmological scenarios. 
DM particles decouple earlier than the standard case, which caused the larger relic
density compared to the standard cosmology. After including 
coannihilation, the relic abundance is decreased due to the
coannihilation. The extent of the decrease depends on the size of
modification.
When the modification factor is large, the increase of the DM relic density will be
sizable. Therefore, after including coannihilation, there is slight decrease of the relic abundance  for larger
modification in contrast with the minor modification.
The same result is obtained for the brane world cosmology in panel (b). The
effect of coannihilation is less sizable for smaller $M_5$.

\begin{figure}[H] 
	\begin{center}
		\hspace*{-0.5cm} \includegraphics*[width=8cm]{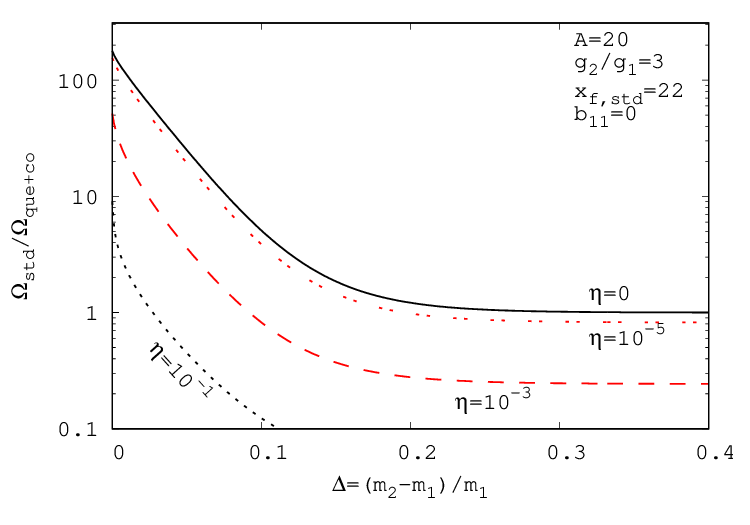}
		\put(-115,-12){(a)}
		\vspace{0.5cm}
		\hspace*{-0.5cm} \includegraphics*[width=8cm]{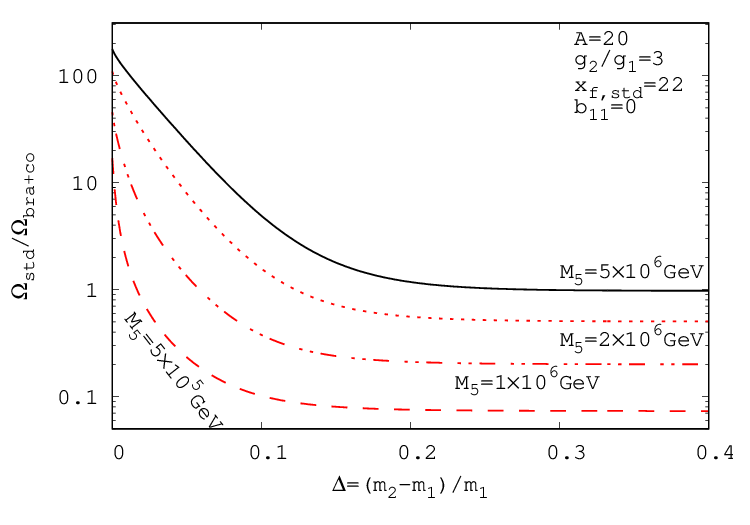}
		\put(-115,-12){(b)}
		\caption{\label{fig:b} \footnotesize
			The ratio of the relic density of DM as a function of
                        relative mass spliting $\Delta$ in standard cosmology including only neutralino pair annihilation $\sigma_{11}$ to the 
			kination model and brane world cosmology including  
			coannihilation $\sigma_{\rm eff}$.    
			Here $m_1 = 100$ GeV, $g_{1} = 2$, $g_* = 90$.}  
	\end{center}
\end{figure}

\begin{figure}[H]
	\begin{center}
		\hspace*{-0.5cm} \includegraphics*[width=8cm]{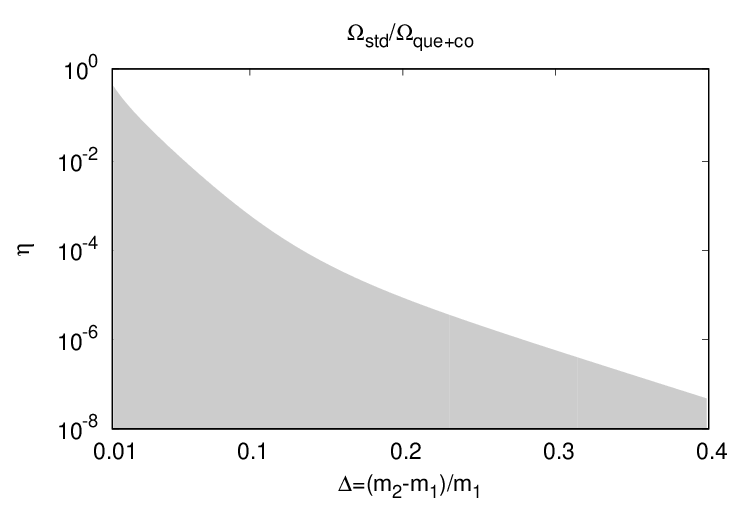}
		\put(-115,-12){(a)}
		\vspace{0.5cm}
		\hspace*{-0.5cm} \includegraphics*[width=8cm]{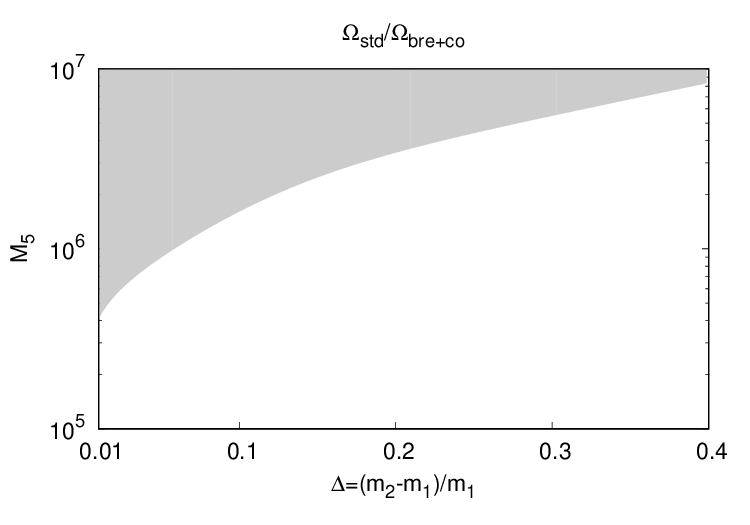}
		\put(-115,-12){(b)}
		\caption{\label{fig:c} \footnotesize
			Contour plots of the relative mass splitting $\Delta$ and $\eta$ ($M_5$)
			when $\Omega_{\rm std}/\Omega_{\rm nstd+co} > 1$.
			Here $A = 20$, $m_1 = 100$ GeV, $g_{1} = 2$, $g_* = 90$. $x_{f,\,\rm std}=22$.}  
	\end{center}
\end{figure}

In Fig.\ref{fig:b}, we plot the ratio of the relic density
in the standard cosmological scenario which only takes into
  neutralino pair annihilation to the non-standard cosmology
which includes coannihilation as a function of the relative mass
splitting $\Delta$. Here, we analyze the difference of the relic abundance
resulted by the combined effect of non-standard expansion and coannihilation.
The decrease of relic density of DM because of coannihilation is quite mild in non-standard cosmology due
to the enhancement of Hubble rate. After including the coannihilation in
non-standard cosmology, the decrease of the relic density is slower than the
standard cosmology.
We take an example of $\eta = 10^{-3}$ in plot ($a$), when the relative mass
splitting $\Delta$ is 0.04, the ratio of the relic density 
in the standard cosmology is 35.7 and in kination model is
5.2. It means there is
already sizable decrease of the relic density with coannihilation in the
standard cosmolgy while the effect starts to be important in kination
model.
The same result is obtained for the brane world cosmology in panel (b). We also noticed that
for larger modification, the effect of coannihilation is insignificant in
Fig.\ref{fig:b}. 
Fig.\ref{fig:c} shows the constraints on $\eta$, $M_5$ and $\Delta$ for which 
the coannihilation to be important. In the shaded regions in panel $(a)$ and
$(b)$, $\Omega_{\rm std}/\Omega_{\rm nstd+co} > 1$. It means in those shaded
regions the coannihilations let the DM relic density decrease.

%
% Fig. 3

% Fig. 4
\begin{figure}[H]
	\begin{center}
		\hspace*{-0.5cm} \includegraphics*[width=8cm]{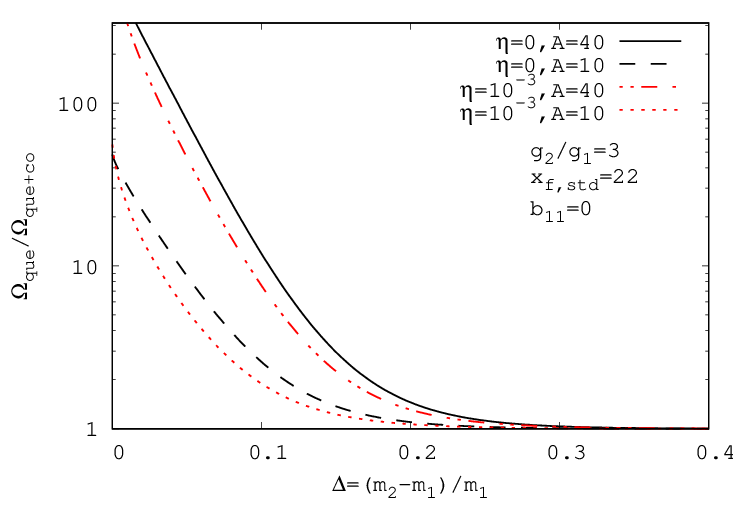}
		\put(-115,-12){(a)}
		\vspace{0.5cm}
		\hspace*{-0.5cm} \includegraphics*[width=8cm]{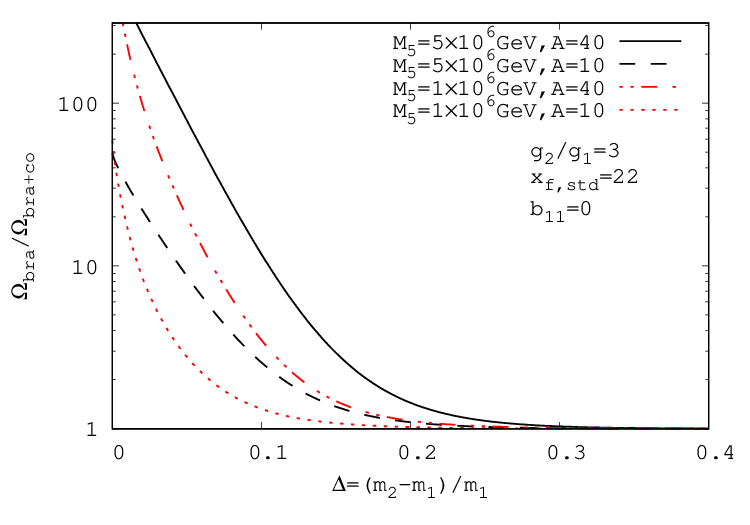}
		\put(-115,-12){(b)}
		\caption{\label{fig:d} \footnotesize
			The ratio of the relic density of DM in non-standard cosmology without
			coannihilation to the case including coannihilation as a function of the
			relative mass splitting $\Delta$ for different cross section enhancements $A$.
			Here $m_1 = 100$ GeV, $g_{1} = 2$, $g_* = 90$.}  
	\end{center}
\end{figure}
The ratio of the relic density of DM in non-standard
cosmology without coannihilation to the case including coannihilation as a
function of the relative mass splitting $\Delta$ for different cross section
enhancements $A$ is plotted in Fig.\ref{fig:d}. We note that the limit for
relative mass splitting for which the coannihilation to be significant in
non-standard cosmology is 
smaller than the standard one. As long as $\Delta < 0.1$ for $A =10$,
there is
sizable effect of coannihilation on the relic abundance in standard cosmology
while the limit becomes $\Delta < 0.08$ in kination model for $\eta = 10^{-3}$
and $\Delta < 0.05$ in brane world cosmology for $M_5=1 \times 10^6$ GeV.
It means when the second lightest particle's mass difference between the
lightest particle's mass is within about
$8\%$ for kination and $5\%$ for brane world cosmology, the
coannihilation becomes relevant. From that figures we can determine how close
in mass $\chi_2$
must be for the coannihilation effect to be important. The decrease of the
relic density is larger for the larger enhancement factor A in kination
and brane world model. The decrease of relic density is also
slower in non-standard cosmology for different $A$ in comparison with the
standard one. The range of relative mass splitting for which the
coannihilation to be important is smaller in those models than in the standard
cosmology.  
For example when A=40, the coannihilation effect is sizable within mass
differences of $15\% $ for standard cosmology and $14\%$ for kination,
$11\%$ for brane world cosmology in Fig.\ref{fig:d}. The enhanced expansion
rate always makes the effect of coannihilation to be weak in different extents.
%

% Fig. 4
\begin{figure}[H]
	\begin{center}
		\hspace*{-0.5cm} \includegraphics*[width=8cm]{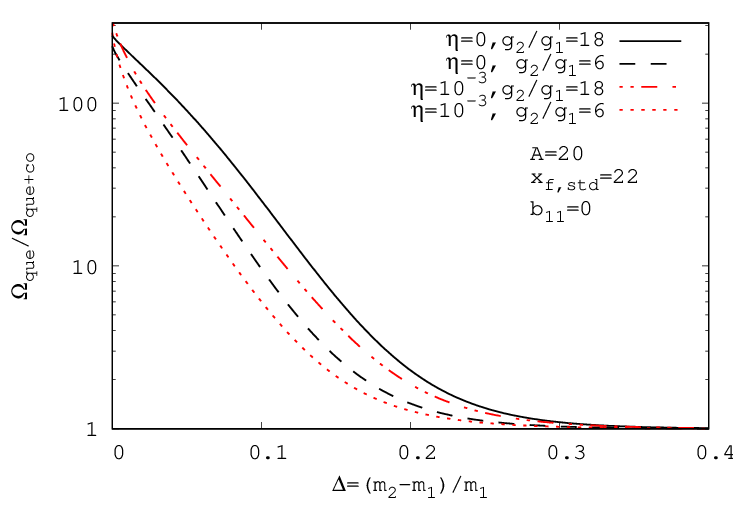}
		\put(-115,-12){(a)}
		\vspace{0.5cm}
		\hspace*{-0.5cm} \includegraphics*[width=8cm]{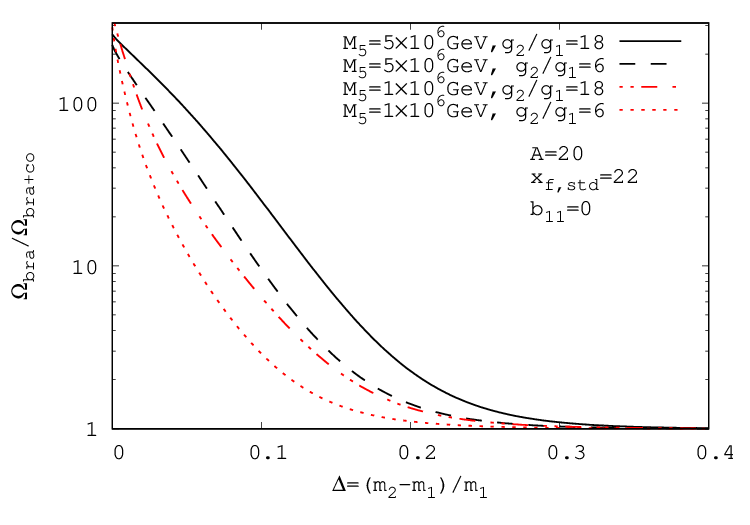}
		\put(-115,-12){(b)}  
		\caption{\label{fig:e} \footnotesize
			$\Omega_{\rm nstd}/\Omega_{\rm nstd+co}$ as a function of the relative mass splitting
			$\Delta$ for different $g_2/g_1$. Here $m_1 = 100$ GeV, $g_{1} = 2$,
			$g_* = 90$.}  
	\end{center}
\end{figure}

Fig.\ref{fig:e} is for $\Omega_{\rm nstd}/\Omega_{\rm nstd+co}$ as a
function of relative mass splitting $\Delta$ when $g_2/g_1$ takes different
values. The effect of coannihilation is  
more relevant when $g_2/g_1$ is larger. On the other hand, when the extent of
the modification is larger in kination and brane world cosmology, the
effect of coannihilation is weaker.

% Fig. 5
\begin{figure}[H]
	\begin{center}
		\hspace*{-0.5cm} \includegraphics*[width=8cm]{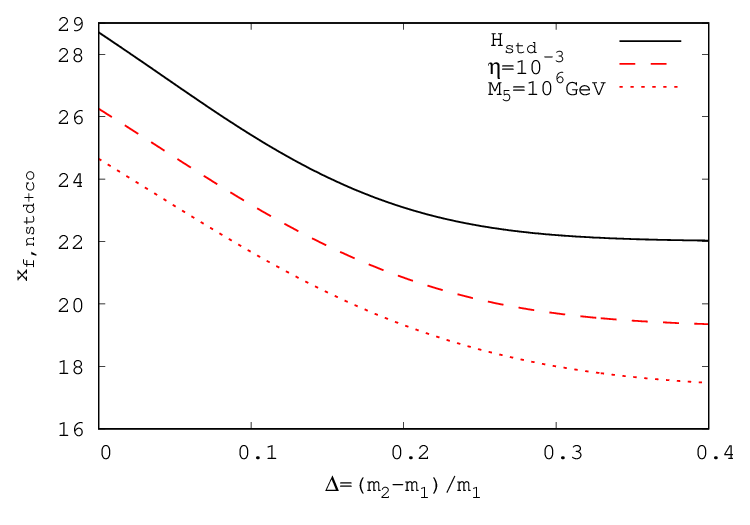}
		\put(-115,-12){(a)}
		\vspace{0.5cm}
		\caption{\label{fig:f} \footnotesize
			The scaled freeze out temperature in non-standard cosmology including
			coannihilation as a function of the relative mass splitting $\Delta$. Here
			$m_1 = 100$ GeV, $g_{1} = 2$, $g_* = 90$, $x_{f,\,\rm std }=22$.}  
	\end{center}
\end{figure}

The scaled freeze out temperature in non-standard cosmology including
coannihilation as a function of the relative mass
splitting is shown in Fig.\ref{fig:f}. The scaled freeze out temperature
is decreased in non-standard cosmology including coannihilation with respect
to the case of stadard one.

\section{Conclusion}
We discussed the coannhilation effect on the relic density of DM in the
quintessence model with kination phase and brane world cosmology. The Hubble
expansion rate is modified in non-standard cosmological scenarios. It
leaves its imprint on the relic density of DM particle. There is increased
relic density of DM due to the enhanced Hubble expansion rate. On the other
hand, the coannihilation mechanism reduces the DM relic density in the standard
cosmological scenario. We found the coannihilations also decrease the relic
density of DM in non-standard cosmological scenarios, while the  enhanced
expansion rate always makes the coannihilation effect to be weak in
various extents. The reduction of relic density because of coannihilation
depends on the size of the modification in non-standard cosmological
scenarios. The decrease is mild for larger modification.

If compared the relic density of DM in standard cosmology including only LSP
pair annihilation to the case of
non-standard cosmology with coannihilation, we note the coannihilation effect is
insignificant for larger modification. The reason is that the relic
density is enhanced sizably when the modification is large in kination
and brane world models. We found constraints on the
modification and relative mass splitting for which the coannihilation to be
relevant. The effect of coannihilation on the relic density is also more
relevant for the larger cross section enhancement $A$ and $g_2/g_1$.

Our result is important to know what extent of the coannihilating particle's
mass should be in order to the coannihilation effect is sizable in the
non-standard cosmological scenarios.

\section*{Acknowledgments}

The work is supported by the National Natural Science Foundation of China
(2020640017, 11765021, 2022D01C52).

\end{document}